\documentclass[%
 reprint,
 amsmath,amssymb,
 aps,
prb,
]{revtex4-2}

\usepackage{graphicx}
\usepackage{dcolumn}
\usepackage{bm}
\usepackage{mathtools} 
\usepackage{braket}
\usepackage{hyperref}
\usepackage{xcolor}

\def\changes{}
\newenvironment{changesenv}
  {}
  {}
\begin{document}

\preprint{APS/123-QED}

\title{Tensor network methods for bound electron-hole complexes beyond strong and weak confinement in nanoplatelets}

\author{Bruno Hausmann}
 \email{b.hausmann@itp.tu-berlin.de}
 \author{Marten Richter}
 \email{marten.richter@tu-berlin.de}
\affiliation{%
 Institut für Physik und Astronomie, Nichtlineare Optik und Quantenelektronik, Technische Universität Berlin, EW 7-1, 10623 Berlin, Germany
}%

\date{\today}

\begin{abstract}
In semiconductor nanostructures, optical excitation typically creates bound electron-hole states, such as excitons, trions, and larger complexes. Their relative motion is described by the Wannier equation, which is valid only for spatially extended motion in the Coulomb-dominated, weak-confinement limit. Other small nanostructures, such as quantum dots, are in the confinement-dominated strong confinement regime, where the wavefunction factorizes into independent electron and hole parts. Nanoplatelets are in between the two regimes and require solving an unfactorized higher-dimensional Schrödinger equation, which is computationally expensive. This work demonstrates how tensor networks can partially overcome this problem, using CdSe nanoplatelets as an example.  The method is also applicable to related two-dimensional systems. As a demonstration, we calculate the excitonic and trionic ground states, as well as several excited states, for nanoplatelets of varying sizes, including their energies and oscillator strengths. More importantly, overall strategies for using tensor networks in real space for systems under intermediate confinement have been developed.
\end{abstract}

\maketitle

\newcommand{\npsize}[2]{$#1\,\text{nm}\times#2\,\text{nm}$}
\section{\label{sec:introduction}Introduction}
Nanoplatelets are thin, boxlike, chemically grown nanostructures made of semiconducting materials (often CdSe) \cite{pelton_carrier_2012, tessier_spectroscopy_2012, benchamekh_tight-binding_2014, pelzer_exciton_2015, naeem_giant_2015, achtstein_p-state_2016, peng_bright_2020, ayari_tuning_2020, antolinez_trion_2020, vong_origin_2021, riesner_demystifying_2024, swift_controlling_2024}. They are a few monolayers thick in the $z$-direction but are significantly larger in the $x$- and $ y$-directions. Because of their two-dimensional nature, nanoplatelets are often promoted as chemically grown alternatives to epitaxially grown quantum wells \cite{ithurria_quasi_2008, pelton_carrier_2012, achtstein_electronic_2012, tessier_spectroscopy_2012}. However, many nanoplatelets are too small perpendicular to the growth direction to be in the Coulomb-dominated weak confinement regime of quantum wells. Therefore, the electron-hole states should not be calculated in relative and center-of-mass (COM) coordinates, as in the quantum-well case.
Instead of the relative coordinate Wannier equation for excitons and its analogue for trions, the full unfactorized Schrödinger equation with its four-dimensional (exciton) and six-dimensional (trion) wave functions needs to be solved. \cite{richter_nanoplatelets_2017}  The high dimensionality makes a direct solution of the Wannier equation computationally expensive and even unfeasible in the case of trions.

Tensor networks have been successfully used to approximate and efficiently represent high-dimensional wave functions (which, in principle, are also high-dimensional tensors). Initially, tensor networks were applied to systems such as spin chains \cite{verstraete_matrix_2006, vidal_classical_2007, clark_exact_2010, cirac_matrix_2017}, where each physical site corresponds to a site index of the tensor network. Quantics tensor trains (QTTs), a type of tensor network, were introduced to solve partial differential equations, \cite{oseledets_approximation_2009, oseledets_approximation_2010, khoromskij_odlog_2011, kazeev_low-rank_2012, oseledets_solution_2012, khoromskaia_tensor_2015, benner_fast_2017} where every tensor index corresponds to a bit of the coordinate. Kuhn and Richter \cite{kuhn_combined_2019, kuhn_tensor_2020} used QTTs to compute trion and biexciton states in unconfined 2d monolayers of transition metal dichalcogenides (TMDCs) on the Brillouin zone of quasimomentum space. QTTs have also been applied to computational fluid dynamics, including turbulence modeling. \cite{gourianov_quantum_2022, kiffner_tensor_2023}

In the following section the model system and its equations are introduced, then in section \ref{sec:methods} the methods utilized and developed for this work are presented: it discusses how tensors arising from the discretization of functions is represented as quantic tensor trains and how the system's Hamiltonian is constructed in this representation; furthermore, some details of the eigenstate calculation using DMRG are discussed. Section \ref{sec:results} presents our results and discusses the calculation of the observables from the QTT states.

\section{Model system and Wannier equation}
Using the electron-hole picture, the crystal Hamiltonian without confinement potential reads, \cite{haken_quantum_1988,haug_quantum_2009}
\begin{equation}
    \begin{split}
        H = &W_{full} \\
        &- \sum_{\mathbf{k}}\varepsilon_{v\mathbf{k}}h^\dag_\mathbf{k}h_\mathbf{k} + \sum_{\mathbf{k}}\varepsilon_{c\mathbf{k}}e^\dag_\mathbf{k}e_\mathbf{k}  \\
        &+ \frac{1}{2}\sum_{\mathbf{k}_1\mathbf{k}_2\mathbf{k}_3\mathbf{k}_4}V^{cccc}_{\mathbf{k}_1\mathbf{k}_2\mathbf{k}_3\mathbf{k}_4}e_{\mathbf{k}_1}^\dag e_{\mathbf{k}_2}^\dag e_{\mathbf{k}_3} e_{\mathbf{k}_4} \\
        &+ \frac{1}{2}\sum_{\mathbf{k}_1\mathbf{k}_2\mathbf{k}_3\mathbf{k}_4}V^{vvvv}_{\mathbf{k}_1\mathbf{k}_2\mathbf{k}_3\mathbf{k}_4}h_{\mathbf{k}_3}^\dag h_{\mathbf{k}_4}^\dag h_{\mathbf{k}_1} h_{\mathbf{k}_2} \\
        &- \sum_{\mathbf{k}_1\mathbf{k}_2\mathbf{k}_3\mathbf{k}_4}V_{\mathbf{k}_1\mathbf{k}_2\mathbf{k}_3\mathbf{k}_4}^{cvvc}e_{\mathbf{k}_1}^\dag h_{\mathbf{k}_3}^\dag h_{\mathbf{k}_2} e_{\mathbf{k}_4} \\
        &+ \sum_{\mathbf{k}_1\mathbf{k}_2\mathbf{k}_3\mathbf{k}_4}V_{\mathbf{k}_2\mathbf{k}_1\mathbf{k}_3\mathbf{k}_4}^{vcvc}e_{\mathbf{k}_1}^\dag h_{\mathbf{k}_3}^\dag h_{\mathbf{k}_2} e_{\mathbf{k}_4},
    \end{split}
\end{equation}
where $e^\dag$ is the creation operator of a conduction band electron, $h^\dag$ is the creation operator for a valence band hole, $W_{full}$ is the energy of the completely filled valence band and $V^{\lambda_1\lambda_2\lambda_3\lambda_4}_{\mathbf{k}_1\mathbf{k}_2\mathbf{k}_3\mathbf{k}_4}$ is the Coulomb matrix element. The second and third terms represent the energy of the electrons and holes, and the following terms represent the different kinds of two-particle interactions: electron-electron, hole-hole, and electron-hole Coulomb repulsion and attraction, as well as electron-hole exchange interaction. \cite{haken_quantum_1988} 
The Coulomb attraction \changes{$V_{\mathbf{k}_1\mathbf{k}_2\mathbf{k}_3\mathbf{k}_4}^{cvvc}$ (sixth term)} can lead to the creation of bound electron-hole complexes, such as an exciton in the case of a single electron and hole. 

Assuming that only states near the band edge at the $\Gamma$-point $\mathbf{k}\approx0$ have a relevant contribution to the bound electron-hole, the full state $\ket{\Psi_E}$ of an exciton for example is described by an envelope wave function $\psi_E(\mathbf{r}_e, \mathbf{r}_h)$ and Bloch functions $u_\lambda(\mathbf{r}_\lambda)$ at $\mathbf{k}\approx0$.
Because of the nanoplatelets' small thickness, electron and hole motion is overall two-dimensional.
This leads to a factorization of the envelope function into the in-plane part $\psi_E(\bm{\rho}_e, \bm{\rho}_h)$ with two-dimensional coordinates $\bm{\rho}_e$, $\bm{\rho}_h,$ and the envelope functions $\zeta_\lambda(z_\lambda)$ for the $z$-direction.
This ansatz results in an equation for the in-plane envelope function:
\begin{eqnarray}
    \left[ -\frac{\hbar^2}{2m_e}\Delta_{\bm{\rho}_e}-\frac{\hbar^2}{2m_h}\Delta_{\bm{\rho}_h}-U(\bm{\rho}_e-\bm{\rho}_h) \right]\psi_E(\bm{\rho}_e, \bm{\rho}_h)= \nonumber \\
    =E_E\psi_E(\bm{\rho}_e, \bm{\rho}_h),\hspace{10pt}
    \label{eq:exciton_wannier_equation}
\end{eqnarray}
where $m_e$ and $m_h$ are the effective masses of the electron and hole, respectively, and $U(\bm{\rho}_e-\bm{\rho}_h)$ is the Coulomb potential between the two particles. \cite{richter_nanoplatelets_2017, ayari_tuning_2020}

The corresponding equation for trions is very similar \cite{ayari_tuning_2020}:
\begin{equation}
    \begin{split}
        &\biggl[-\frac{\hbar^2}{2m_e}\Delta_{\bm{\rho}_{e1}}-\frac{\hbar^2}{2m_e}\Delta_{\bm{\rho}_{e2}}-\frac{\hbar^2}{2m_h}\Delta_{\bm{\rho}_h} \\
        &\quad-U(\bm{\rho}_{e1}-\bm{\rho}_h)-U(\bm{\rho}_{e2}-\bm{\rho}_h) +U(\bm{\rho}_{e1}-\bm{\rho}_{e2}) \\
        &\quad \biggr] \psi_T(\bm{\rho}_{e1}, \bm{\rho}_{e2}, \bm{\rho}_h)=E_T\psi_T(\bm{\rho}_{e1}, \bm{\rho}_{e2}, \bm{\rho}_h),
    \end{split}
    \label{eq:trion_wannier_equation}
\end{equation}
where $\psi_T(\bm{\rho}_{e1}, \bm{\rho}_{e2}, \bm{\rho}_h)$ is the envelope function describing a negatively charged trion state.

Due to the screening effects of the surrounding dielectrics, the Coulomb potential is modified in two-dimensional materials and is described by the Rytova-Keldysh potential \cite{berkelbach_theory_2013,Keldysh1979}
\begin{equation}
    U_{K}(\rho)=\frac{e^2}{8\pi\epsilon_0\epsilon_{r,out}\rho_0}\left[\ln{\frac{\rho}{\rho+\rho_0}}+\left(\gamma-\ln{2}\right)e^{-\rho/\rho_0}\right],
\end{equation}
where $\rho_0=z_0\epsilon_r/(2\epsilon_{r,out})$, $z_0$ is the platelet thickness, $\epsilon_r$  is the dielectric constant of the platelet, and $\epsilon_{r,out}$ is the dielectric constant of the surrounding solvent. $\gamma$ denotes Euler's constant.
Of course, the Rytova-Keldysh potential assumes an infinitely extended two-dimensional system, which is only approximately true for nanoplatelets.

The logarithmic singularity is hard to treat numerically. Therefore, in Ref. \onlinecite{richter_nanoplatelets_2017}  the approximate potential
\begin{equation}
    U_{Approx}(\rho)=-\frac{e^2}{4\pi\epsilon_0\epsilon_r}\frac{1}{\sqrt{\rho^2+(\alpha_0z_0)^2}}
    \label{eq:approx_keldysh_potential}
\end{equation}
was used analogous to \cite{mayrock_weak_1999} , which does not have a singularity. The parameter $\alpha$ is tuned to reproduce the correct energy for the lowest energy eigenstates of an exciton in an infinitely extending platelet. \cite{richter_nanoplatelets_2017}

\section{\label{sec:methods}Methods}
\subsection{Representation of function discretizations using quantics tensor trains}
Tensor networks can serve as compressed representations of high-rank, high-dimensional tensors that are too large to store in memory.
Matrix product states (MPS) are a tensor network for which many numerical methods for optimization \cite{holtz_alternating_2011}, eigenvector calculation (DMRG \cite{white_density_1992,ostlund_thermodynamic_1995,dukelsky_equivalence_1998,white_density_2005}, imaginary time evolution \cite{schollwock_density-matrix_2011}, and others \cite{baiardi_excited-state_2022}), and temporal propagation exist \cite{vidal_efficient_2004, daley_time-dependent_2004, haegeman_time-dependent_2011, haegeman_unifying_2016, paeckel_time-evolution_2019}. The key to the efficiency of these methods is that they operate directly on the MPS format, rather than requiring it to be uncompressed (contracted) and recompressed for every operation.
As a result, MPS is among the most widely used tensor network types. In an MPS, a high-dimensional tensor $A_{s_1,s_2,...s_N}$ of rank N is represented as
\begin{equation}
    A_{s_1,s_2,...,s_N}=\sum_{l_1,l_2,...,l_{N-1}}B^{(1)}_{s_1,l_1}B^{(2)}_{l_1,s_2,l_2}...B^{(N)}_{l_{N-1},s_N},
    \label{eq:mps_decomposition}
\end{equation}
all tensors $B^{(i)}$ in the MPS are only of rank 3 (or rank 2 for $B^{(1)}$ and $B^{(N)}$). This representation is possible for any tensor $A_{s_1,s_2,...s_N}$, but it is only efficient when the dimensions of the link indices $l_1$,$l_2$,...,$l_{N-1}$ required for an accurate representation are not too big. The original tensor $A$ requires storing $d^N$ values, where $d$ is the dimension of the site indices $s_1,s_2,...s_N$, while the MPS representation requires $O(dm^2N)$ values, where $m$ is the maximum link dimension. \cite{vidal_efficient_2003, schollwock_density-matrix_2011}

Linear operators on MPS are represented in the MPO format, which is similar to MPS but has one input and one output index per site.

We discretize a function $f(x)$ of a continuous variable $x$ on a regular grid $x_i=i\Delta x$ with the grid discretization $\Delta$ resulting in the tensor $f_i=f(x_i)$. The resulting tensor can often be efficiently represented as a QTT \cite{oseledets_approximation_2009, oseledets_approximation_2010, khoromskij_odlog_2011, dolgov_fast_2012, oseledets_solution_2012, khoromskaia_tensor_2015, benner_fast_2017}. To construct a QTT from the tensor $f_i$, the index $i$ is replaced by the bits $\sigma^1,...,\sigma^N$ of its binary representation:
\begin{equation}
    i=\sum_{i_{bit}=1}^N\sigma^{i_{bit}}2^{i_{bit}-1}.
	\label{eq:i_to_sigma}
\end{equation}
The resulting tensor of rank $N$ is then decomposed into the format of an MPS, this result is called a QTT. Again the efficiency of the QTT representation depends on the maximum link dimension. Ideally, the required maximum link dimension remains constant as resolution $K$ increases, so that the required storage scales logarithmically: $O(2^m\log_2 K)$.
 
A QTT representation can also be obtained for functions of multiple variables $f(x, y, ...)$ by discretizing each variable separately with a set of bits $\sigma^1_i,...,\sigma^N_i$ for each variable. The order in which these bits are arranged into the site indices during decomposition is crucial for efficiency. In the multivariate case, one might try to construct an MPS with the original indices for $x,y,...$ as the site indices instead of their bits, but this would result in large site dimensions and thus large site tensors. The crucial point of the QTT format is this reduction of the site dimension\changes{, which will result in a more efficient representation, provided that the required link dimensions do not become too big. For MPS that represent physical systems and have a correspondence between MPS sites and physical sites (such as a spin chain), the link dimension depends on the entanglement of the sites. In principle, this can be transferred to the QTTs, where the QTT/MPS sites no longer represent physical sites but instead bits, each bit is associated with a length scale: bits with higher weight in the binary representation (Eq. (\ref{eq:i_to_sigma})) correspond to larger length scales, while bits with smaller weight correspond to smaller length scales. The efficiency of the QTT approach should thus depend on the entanglement or correlation between the length scales.}

The following two subsections explain how the terms of the Hamiltonian, from the governing equations \ref{eq:exciton_wannier_equation} and \ref{eq:trion_wannier_equation}, are expressed as MPOs that act on QTTs.

\subsection{Construction of finite difference operators for QTTs}
Partial derivatives of functions discretized on a grid can easily be approximated using finite differences.
The second derivative that is needed for the conversion of the Laplace operator $\Delta$ in Eqs. (\ref{eq:exciton_wannier_equation}) and (\ref{eq:trion_wannier_equation}) becomes
$f''_i=f''(i\Delta x)\approx\left(f_{i+1}+f_{i-1}-2f_{i}\right)/\Delta x^2,$
approximated to second order. We will use only the lowest order finite difference approximation to keep the complexity and, thus, the overall link dimension of the MPO low. To achieve higher accuracy, we would rather increase the number of grid points (and thus bits) than use a higher-order finite difference approximation.
Generally, finite differences approximate a derivative at grid point $i$ as a linear combination of the function value at $i$ itself and the neighboring points $i+1,i-1,i+2,i-2,...$.

\begin{figure}[t]
\includegraphics[width=\linewidth]{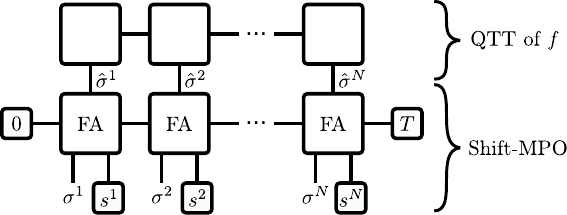}
\caption{\label{fig:shift_mpo} Shift operator applied to the QTT representation of $f$ consisting of the addition network (full adder tensors FA), the one-hot tensors representing the shift $s$, and the termination tensors on both ends of the addition network.}
\end{figure}

Therefore, the first step in constructing the MPO representation of finite difference operators would be to construct a grid shift operator that transforms the QTT corresponding to the values $f_i$ into the QTT corresponding to $f_{i+s}$, which is shifted by $s$ grid points. A finite difference MPO can then be built by linearly combining (adding) multiple shift operators with different shifts $s$.

Such a shift operator is implemented using logical circuits \cite{kuhn_combined_2019, kuhn_tensor_2020}. The shift MPO has to perform a binary addition of its input indices and the index shift $s$ to obtain the bits of the shifted index $i+s$. A binary addition is implemented using the tensor representation of an addition network, a logical circuit consisting of a chain of full adders. Each full adder has three inputs: one for the previous carry bit, two for the bits to be added, and two outputs: the next carry bit and the result at this position. Fig. \ref{fig:shift_mpo} illustrates the final construction applied to the QTT of $f$.

This logical circuit is translated to an MPO by the following procedure:
\begin{enumerate}
    \item Each logical circuit $f: \{0, 1\}^{N_{in}}\rightarrow\{0, 1\}^{N_{out}}$ (i.e., each full adder)\changes{, where $N_{in}$ is the number of input bits and $N_{out}$ is the number of output bits,} becomes a tensor with one index for each input and output. Each combination of input values $\changes{(i_1,i_2,...,i_{N_{in}})}$ is linked with its output values $\changes{(o_1,o_2...,o_{N_{out}})}$ by setting the corresponding tensor element to one, while the other elements remain zero: 
    \[\changes{f^{i_1,i_2...,i_{N_{in}}}_{o_1,o_2...,o_{N_{out}}}} = \left\{
\begin{array}{lll}
    1 & \changes{f(i_1,i_2...,i_{N_{in}})} \\
      & =\changes{(o_1,o_2...,o_{N_{out}})} \\
    0 & \text{otherwise}
\end{array}
\right..\] 
    Thus, only the input and output values appearing in the logical circuits' truth table contribute to the sum when the tensor is contracted with other tensors.
    \item Where the input and output of two logical elements are connected, the corresponding tensors are contracted along their respective indices.
\end{enumerate}

To build a shift MPO from the converted full-adder circuit, the summand bits corresponding to shift $s$ in the tensor network must be fixed. This is achieved by multiplying the respective half-adder input indices with one-hot tensors (tensors where all entries except one with value one are zero), which encode the shift bits, these tensors are labeled $s^1$ to $s^N$ in Fig. \ref{fig:shift_mpo}. The other set of input indices becomes the input indices of the shift MPO, $\sigma^1$ to $\sigma^N$ in the figure, whereas the full adder output indices become the output indices of the MPO, $\hat{\sigma}^1$ to $\hat{\sigma}^N$.

The termination of the last carry index of the adder circuits directly sets boundary conditions. Because if the addition of the shift operator leads to a value outside the possible index range $K$, the carry bit of the last full adder outputs $c_{out}=1$. If we impose Dirichlet boundary conditions, the result of such a shift should be zero, corresponding to ghost points outside the domain with zero values. This is achieved by multiplying the last carry index by another one-hot tensor, the termination tensor $T_{c_{out}}$, which is 1 only for $c_{out}=0$ and 0 otherwise. 

If, instead, we choose for termination tensor $T_{c_{out}}=1$ $\forall c_{out}$, then the overflow of binary addition results in periodic boundary conditions.\\
The first carry index, which corresponds to the input carry of the first full adder, is always set to zero by multiplying it by a one-hot tensor encoding zero (labeled 0 in Fig. \ref{fig:shift_mpo}).
Negative shifts are implemented using two's complement representation, in which case an additional dummy bit must be added.

\subsection{Construction of the two-particle potential as a QTT}
The two-particle Coulomb potentials are implemented similarly using logical circuits. 
The two-particle potential $U(\mathbf{r}_1-\mathbf{r}_2)$ only depends on the difference of the positions of both particles $\mathbf{r}_{12}=\mathbf{r}_1-\mathbf{r}_2$. Therefore, we first discretize $U(\mathbf{r}_{12})$ and decompose the resulting tensor into an effectively single-particle QTT, and then construct the full two-particle potential QTT from the resulting tensor in a second step .

\begin{figure}[b]
\includegraphics[width=\linewidth]{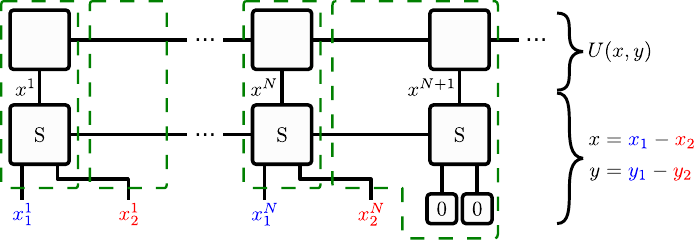}
\caption{\label{fig:two_particle_potential} QTT representation of a two-particle potential built from the effective single particle potential $U(x, y)$ (top) and a subtraction network (bottom). The tensors that are contracted into a single tensor in the QTT are indicated by the dotted contour boxes.}
\end{figure}

Then a subtraction network is attached to the single-particle QTT, which computes $\mathbf{r}_{12}=\mathbf{r}_1-\mathbf{r}_2$ on the level of the bit indices (cf. Fig. \ref{fig:two_particle_potential}.) Similar to the addition network, the subtraction network consists of a chain of subtractors, each taking in two inputs: one for each particle coordinate. \footnote{Since $a+b=c\Leftrightarrow a=c-b$ a subtraction network can be implemented simply by exchanging the indices of an addition network.} These become the new indices of the two-particle QTT. The subtractor outputs are connected with the old indices of the single-particle potential QTT.

The grid used to discretize $U(\mathbf{r}_{12})$ also contains negative values of $\mathbf{r}_{12}$ and is twice as big in each dimension as the grid for $\mathbf{r}_1$ or $\mathbf{r}_2$. Thus, each dimension requires an additional bit, and two's complement is used to encode negative values. 
The additional bit is connected to the output of the last subtractor; its inputs are both set to zero.

To improve the network's representation efficiency, a variational compression algorithm \cite{schollwock_density-matrix_2011} is applied in the final step.

\subsection{Control of DMRG sweeps and convergence criterion}
\label{sec:control_protocol}
With the Hamiltonian expressed as a sum of MPOs, the stationary Schrödinger Eqs. (\ref{eq:exciton_wannier_equation}) and (\ref{eq:trion_wannier_equation}) are solved using the well-established DMRG \cite{white_density_1992,ostlund_thermodynamic_1995,dukelsky_equivalence_1998,white_density_2005} algorithm. DMRG takes a Hermitian operator (the Hamiltonian), represented as an MPO or as a sum of MPOs, and finds the ground state in MPS form via variational optimization. As we will discuss in section \ref{sec:dmrg_excited_states}, DMRG can also be adapted to find excited states. 

During the variational optimization, DMRG repeatedly sweeps over the MPS. In two-site DMRG \cite{schollwock_density-matrix_2011}, the links can adaptively grow and shrink during the sweep. This is typically controlled by a maximum allowed link dimension or a cutoff value, and these parameters for each sweep are crucial to the efficiency of the DMRG algorithm. Too-aggressive truncations lead to DMRG failure, and too-low truncation leads to a slow, memory-intensive, and infeasible algorithm. 

Therefore, a control protocol was used that starts with aggressive truncation and a random initial state with small link dimension, \changes{after the energy has converged at this truncation level and changes only minimally from one sweep to another, the truncation is relaxed (increase of the maximum link dimension/decrease of the cutoff) and the DMRG sweeps are continued. This procedure is repeated until the state is deemed close enough to an eigenstate. The aim of this approach is to keep the final state as simple (lowest link dimension) as possible for the desired convergence measure}

A possible option \changes{for the convergence measure} is \changes{a threshold $(\Delta E_{tol})^2$ for} the energy variance \cite{schollwock_density-matrix_2011,hubig_error_2018}
\begin{equation}
    \changes{(\Delta E)^2=\braket{H^2}-E^2\leq(\Delta E_{tol})^2,}
    \label{eq:energy_variance_criterion}
\end{equation}
which is zero only for an eigenstate. We employed this measure in our calculation, but the $\braket{H^2}$ term makes it numerically demanding. Additionally, if implemented using a simple contraction to compute the two expectation values $\braket{H^2}$ and $E=\braket{H}$, the measure can yield incorrect results due to rounding errors. Alternative measures, such as the \textit{2-site-variance} \cite{hubig_error_2018}, might thus be a better choice.

\begin{figure}[t]
\includegraphics[width=\linewidth]{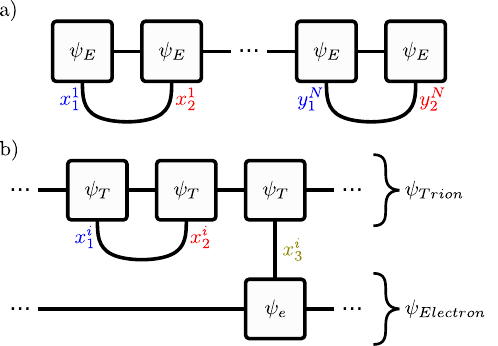}
\caption{\label{fig:trion_oscillator_strength}\label{fig:exciton_osccilator_strength} Contraction scheme for computing the oscillator strengths of: a) the transition from the crystal ground state into the exciton state $\psi_E$  and b) the transition from the single electron state $\psi_{Electron}$ into the trion state $\psi_{Trion}$.}
\end{figure}
\begin{figure}[t]
    \centering
    \includegraphics[width=\linewidth]{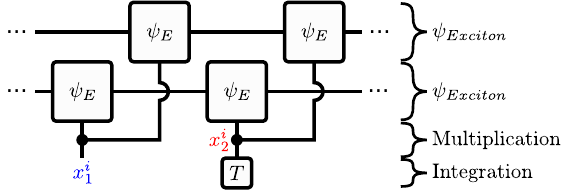}
    \caption{Contraction scheme for computing the electron density $|\tilde{\psi}_e(\mathbf{r}_e)|^2=\int d^2\mathbf{r}_h\left|\psi_E(\mathbf{r}_e, \mathbf{r}_h)\right|^2$ from the QTT representation of $\psi_E$. $T$ denotes a tensor filled with ones that is used to perform a summation (integration) over an index, here they are applied to all indicies belonging to the hole coordinates. Converseley, for the hole density $|\tilde{\psi}_h(\mathbf{r}_h)|^2$ they would be applied to the electron coordinates. If desired $T$ tensors are also applied to the least significant bits to reduce the resolution.}
    \label{fig:electron_density}
\end{figure}

\subsection{\label{sec:dmrg_excited_states} Computation of excited states using DMRG}
The DMRG algorithm retrieves the ground state of a Hermitian operator $H$. A simple way to also compute excited states using DMRG is to iteratively compute higher-energy states by applying weighted projectors onto the Hamiltonian to the already found states, thereby penalizing them. This is already implemented in ITensors.jl \cite{ITensor,ITensor-r0.3} for a Hamiltonian represented by a single MPO. We extended this implementation to handle a sum of MPOs. However, the variational optimization performed by DMRG can sometimes get stuck in a local minimum; in this case, the result is not the ground state but an excited state. Thus, some excited states might be missed when this iterative method is used. Nonetheless, they can still be found in subsequent iterations.

\section{\label{sec:results}Results}
For all numerical calculations, the following parameters were used: electron mass $m_e=0.22m_0$, hole mass $m_h=0.41m_0$\cite{benchamekh_tight-binding_2014}, dielectric constant of the platelet $\epsilon_r=9.5$, dielectric constant of the solvent $\epsilon_{r,out}=5$, potential fit parameter $\alpha_0=1.1$ and platelet thickness $z_0=4.5\cdot0.302\,\text{nm}$\cite{achtstein_electronic_2012}, these are the same as in Ref. \onlinecite{richter_nanoplatelets_2017}. \\
We used the ITensors.jl Julia \cite{ITensor,ITensor-r0.3} library for our implementation.

\changes{Since, we are mainly interested in the development of the QTT based method, the calculations in this section were all performed using the numerically less demanding approximate potential. However, in Appendix \ref{app:full_keldysh_excitons} we compare the results to simulations performed using the full Keldysh potential.} 

\subsection{Excitons}
In Ref. \onlinecite{richter_nanoplatelets_2017}, the full exciton Schrödinger equation (Eq. \ref{eq:exciton_wannier_equation}) was solved directly -- without using tensor networks -- for nanoplatelets of various sizes. These results were used to validate the tensor network methods developed here. Various observables were computed from the states in QTT representation for this comparison. One observable was the oscillator strength defined as $|O|^2\propto\left|\int \text{d}r^2 \psi_{E}(\mathbf{r}, \mathbf{r})\right|^2$ which can easily be calculated from the QTT representation by contracting the QTT tensors as illustrated in Fig. \ref{fig:exciton_osccilator_strength}. The other observables were the electron $|\tilde{\psi}_e(\mathbf{r}_e)|^2=\int d^2\mathbf{r}_h\left|\psi_E(\mathbf{r}_e, \mathbf{r}_h)\right|^2$ and hole $|\tilde{\psi}_h(\mathbf{r}_h)|^2=\int d^2\mathbf{r}_e\left|\psi_E(\mathbf{r}_e, \mathbf{r}_h)\right|^2$ projections of the wave function, which can also be easily computed in the QTT representation (see Fig. \ref{fig:electron_density}):\\ First, the complete density $\left|\psi_E(\mathbf{r}_e, \mathbf{r}_h)\right|^2$ is calculated by multiplying the QTT element-wise with itself. Second, one of the coordinates $r_h$ or $r_e$ is integrated out by contracting the respective site indices with a vector filled with ones (denoted by $T$ in Fig. \ref{fig:electron_density}), which simply means that the index is summed. Additionally, since we are only interested in the densities for visualization, we can also integrate out the least significant bits to obtain a lower resolution. These steps are not executed sequentially. Rather, the resulting network is contracted from left to right in one step.

Ref. \onlinecite{richter_nanoplatelets_2017} also reports the projections $|\tilde{\psi}_{\text{COM}}(\mathbf{R})|^2=\int d^2\mathbf{r} \left|\psi_E(\mathbf{R}, \mathbf{r}) \right|^2$ and 
$|\tilde{\psi}_{\text{r}}(\mathbf{r})|^2=\int d^2\mathbf{R} \left|\psi_E(\mathbf{R}, \mathbf{r}) \right|^2$,  with the center-of-mass $\mathbf{R}=\left(m_e\mathbf{r}_e+m_h\mathbf{r}_h\right)/\left(m_e+m_h\right)$ and the relative coordinate $\mathbf{r}=\mathbf{r}_e-\mathbf{r}_h$. Here, the computation of the relative coordinate is achieved by using a logical circuit similar to the one used in constructing the two-particle potential. However, such a logical circuit cannot easily calculate the center-of-mass coordinate because of the multiplication with the relative masses $m_h$ and $m_e$. 

\begin{figure}[t]
    \centering
    \includegraphics[width=\linewidth]{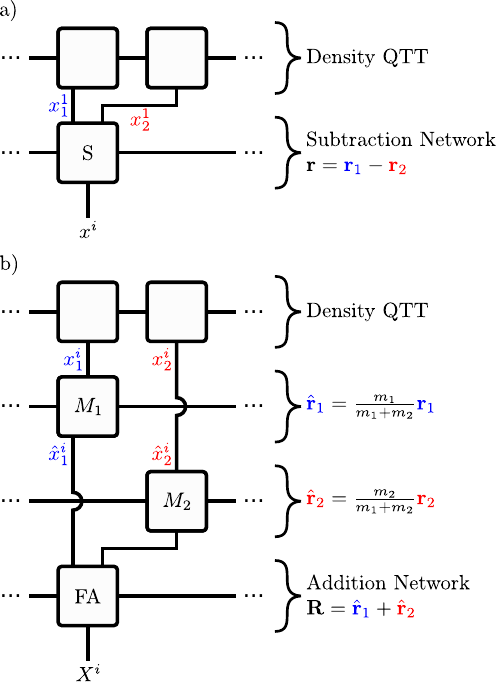}
    \caption{Construction of the a) relative $|\tilde{\psi}_{\text{r}}(\mathbf{r})|^2$ and b) center-of-mass densities $|\tilde{\psi}_{\text{COM}}|^2$ from a QTT representing the original density $\left|\psi_E(\mathbf{r}_e,\mathbf{r}_h)\right|^2$ (top). The layers below the density QTT are applied one after another by interpreting them as MPOs.}
    \label{fig:rel_com_densities_combined}
\end{figure}

This does not apply to the computation of the relative density $|\tilde{\psi}_{\text{r}}(\mathbf{r})|^2$, because here the center-of-mass coordinate is traced out anyway, requiring no actual center-of-mass transformation. The procedure, illustrated in Fig. \ref{fig:rel_com_densities_combined} a), is as follows: First, the QTT representing the wavefunction is multiplied element-wise with itself to obtain a QTT representation of the density $\left|\psi_E(\mathbf{r}_e, \mathbf{r}_h) \right|^2$. Then, one applies a subtraction network for each dimension to obtain the new set of bit indices for $\mathbf{r}$ and to integrate out the other coordinate.

But for the center-of-mass density $|\tilde{\psi}_{\text{COM}}|^2$, the transformation $\mathbf{R}=\left(m_e\mathbf{r}_e+m_h\mathbf{r}_h\right)/\left(m_e+m_h\right)$ is required. The first task is to encode the multiplication of a coordinate $y=m x$ in a linear operator/tensor $M_{ij}$ (to be decomposed into a tensor network later on) with an input index $i$ (corresponding to a grid point $x_i$) and an output index $j$ (corresponding to a grid point $y_i$). Applying the logic behind the previous index transformations, this tensor $M_{ij}$ should be one if the multiplication of the coordinate $x_i$ by $m$ results in the value $y_i$ and zero otherwise. Of course, for an arbitrary output grid $y_1,y_2,...,y_K$, the multiplication result $y$ will generally not match any grid point exactly. One way is to choose a grid adapted to $m$; namely, choosing $y_i=mx_i$ would make the tensor trivial: $A_{ij}=\delta_{ij}$. But, because $m_e\neq m_h$, this would only push the problem into the addition step, where two coordinates from two different grids would need to be added. A working solution is to use the same output grid for both multiplications and to simply round the multiplication result $y$ to the nearest point $y_i$ on the grid. In our implementation, we use bilinear interpolation instead. 
Because the tensor $M_{ij}$ transforms only a single coordinate, it has only $K^2$ values, so it is computed and stored directly in memory. After construction, it is decomposed by repeated SVDs into an MPO acting on the bit indices. This MPO is then applied to the density QTT $\left|\psi_E(\mathbf{r}_e, \mathbf{r}_h) \right|^2$, after this the scaled coordinates are added using an addition network, and one obtains the QTT representing the desired $|\tilde{\psi}_{\text{COM}}|^2$, this combination of MPOs is shown in Fig. \ref{fig:rel_com_densities_combined} b).

All ground states and excited states (Tables \ref{tab:exciton_state_6x4}-\ref{tab:exciton_state_24x20}) agree with the results reported in Ref. \onlinecite{richter_nanoplatelets_2017}. The energies computed with our method are generally slightly higher, likely due to the higher resolution: convergence analyses for all states indicate that the \changes{energy of states converged according to the variance criterion (Eq. (\ref{eq:energy_variance_criterion}))} increases with resolution. (This trend stems from the kinetic energy, whereas the potential energy decreases.) \changes{More details about both the comparison with Ref. \onlinecite{richter_nanoplatelets_2017} and the convergence analyses are included in Appendix \ref{app:conventional_method} and Appendix \ref{app:convergence_analyses} respectively.} Some excited states are missed by the iterative procedure described in subsection \ref{sec:dmrg_excited_states}, but this includes only states close to the end of the computed energy range.

One advantage of the tensor network method is that higher resolutions are attainable; here, the highest resolution was $N=11$ bits per dimension. At this resolution, the conventional method would require an unrealistic $128\,\text{TiB}$ \footnote{Assuming usage of 64-bit floats.} only to store one wave function. In contrast, the method described here required only megabytes to store the final states, and every state was computed in under 10 minutes on a simple consumer processor. Also, the method does not require a specialized basis set, as is often used in quantum chemistry (cf. Gaussian or Slater-type orbitals \cite{jensen_introduction_2007,slater_atomic_1930,boys_electronic_1950}).

Although this is impressively quick, the computation time \changes{required to achieve convergence as measured by the energy variance criterion (Eq. (\ref{eq:energy_variance_criterion}))} still appears to scale exponentially with the number of bits $N$. \changes{This scaling is due to the required number of sweeps increasing exponentially with $N$. The maximum link dimension, obtained through our truncation control protocol, also increases linearly with $N$, but this only leads to a cubic rise in the duration of each DMRG sweep.} Still, exponential scaling in the number of bits $N$ implies power-law $K^a$ scaling in the resolution $K$; we observe an exponent around $a\approx1$, which varies with platelet size and excitation level.

\begin{table}[t]
    \centering
    \includegraphics{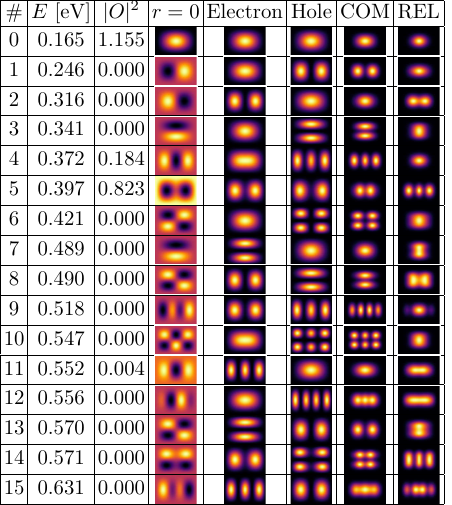}
    \caption{Ground and excited exciton states computed for the \npsize{6}{4} nanoplatelet. The columns list the energy $E$ in eV, oscillator strength $|O|^2$ in arbitrary units (not comparable between platelets), wavefunction $\psi_E\left(\mathbf{R}, \mathbf{r}=0\right)$, and various projected densities: \textit{Electron} for the electron density $|\tilde{\psi}_e(\mathbf{r}_e)|^2$, \textit{Hole}, \textit{COM} and \textit{REL} for the hole $|\tilde{\psi}_h(\mathbf{r}_h)|^2$, center-of-mass $|\tilde{\psi}_\text{COM}(\mathbf{R})|^2$, and relative $|\tilde{\psi}_r(\mathbf{r})|^2$ density.}
    \label{tab:exciton_state_6x4}
\end{table}
\begin{table}[t]
    \centering
    \includegraphics{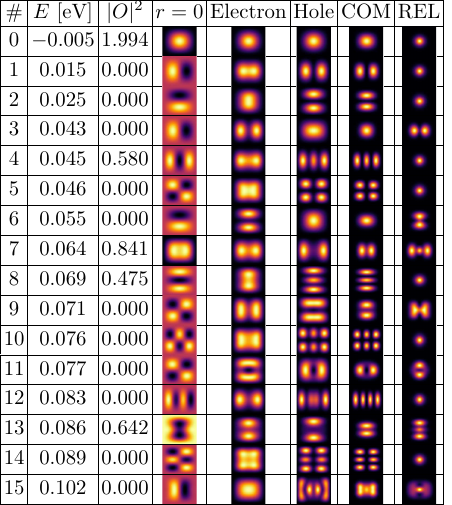}
    \caption{Ground and excited exciton states computed for the \npsize{12}{10} nanoplatelet. See Tab. \ref{tab:exciton_state_6x4} for explanation of the columns.}
    \label{tab:exciton_state_12x10}
\end{table}
\begin{table}[t]
    \centering
    \includegraphics[width=\linewidth]{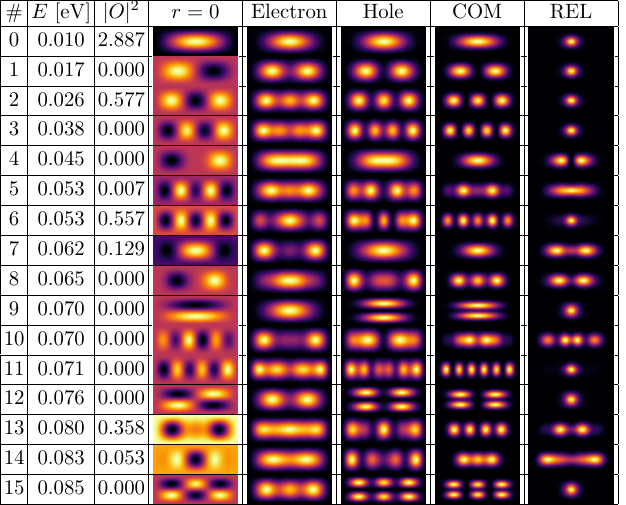}
    \caption{Ground and excited exciton states computed for the \npsize{21}{7} nanoplatelet. See Tab. \ref{tab:exciton_state_6x4} for explanation of the columns.}
    \label{tab:exciton_state_21x7}
\end{table}
\begin{table}[t]
    \centering
    \includegraphics{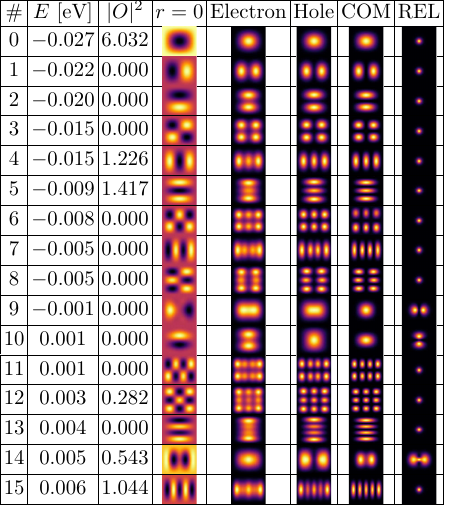}
    \caption{Ground and excited exciton states computed for the \npsize{24}{20} nanoplatelet. See Tab. \ref{tab:exciton_state_6x4} for explanation of the columns.}
    \label{tab:exciton_state_24x20}
\end{table}

\subsection{Trions}
To compute the negatively charged trion eigenstates, the indices in the QTT were arranged as in the exciton case: the $x$- and $y$-coordinates were again partitioned into two separate blocks. Inside each of these blocks, the indices belonging to different particles but to the same bit position were put closest to each other. This pattern is visible in Fig. \ref{fig:trion_oscillator_strength}b), illustrating how the oscillator strength is computed from the QTT representation.

The oscillator strength for the dipole transition from the one-conduction-electron state $\psi_e(\mathbf{r}_e)$ into the trion state described by the spatial wave function $\psi_T(\mathbf{r}_{e1}, \mathbf{r}_{e2}, \mathbf{r}_{h})$ is given by
\begin{equation}
    |O|^2\propto\left|\int \text{d}r^2 \int \text{d}r_e^2 \psi_T^\ast(\mathbf{r}_e, \mathbf{r}, \mathbf{r}) \psi_e(\mathbf{r}_e)\right|^2,
    \label{eq:trion_oscillator_strength}
\end{equation}
which can be easily computed by contracting the QTT of the one-electron ground state with the trion QTT, as illustrated in Fig. \ref{fig:trion_oscillator_strength}b).

Computing trion states is significantly more resource-intensive than computing the exciton states. In practice, at a resolution of $N=11$ bits, computing the trion states was approximately 50 times slower, but this varied widely with platelet size. However, this number compares favorably to the $4.2\cdot10^6$-fold increase in the number of grid points needed to represent the trion wave function compared to the exciton wave function.

\begin{figure}[t]
\includegraphics[width=\linewidth]{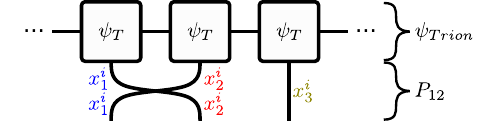}
\caption{\label{fig:permutation_operator} Permutation operator that exchanges the coordinate of the two electrons applied to the trion state $\psi_{Trion}$.}
\end{figure}

When calculating excited states, the DMRG algorithm sometimes struggled to converge when two eigenstates lay energetically close. These eigenstates were often pairs of singlet and triplet states, meaning states where the spatial wave function is symmetric, respectively antisymmetric, under electron coordinate exchange. Therefore, this problem was alleviated by splitting the spectrum computation into two independent parts: one that includes only the triplet states and the other that includes only the singlet states. This separation was achieved by adding a penalty term $\pm wP_{12}$ to the Hamiltonian that penalizes the unwanted symmetry, where $w$ is the penalty energy (always set to $1\,\text{eV}$ in our calculations), $P_{12}$ is the permutation operator (cf. Fig. \ref{fig:permutation_operator}), and the sign is chosen based on the desired symmetry: $+$ to obtain triplets and $-$ for singlets.
Still, larger platelets, whose eigenstates are more densely populated, required significantly more DMRG sweeps to converge. However, it was observed that starting from a low resolution (e.g., $N=4$) and iteratively increasing it by adding one bit per variable until the final resolution ($N=11$) is reached significantly reduces the convergence time.

For all computed states, the oscillator strength, as well as the electron, hole, center-of-mass, and relative densities, were computed and are given in Tables \ref{tab:trion_states_6x4}, \ref{tab:trion_states_21x7}, \ref{tab:trion_states_12x10}, and \ref{tab:trion_states_24x20} for the four platelets. There, the binding energy $E_B$ is defined as the difference between the absolute energy of the trion state and that of the excitonic ground state.\footnote{We calculate $E_B=E_T-E_E$, where $E_T$ is the eigenvalue from Eq. \ref{eq:trion_wannier_equation} for the specific trion state and $E_E$ is the corresponding ground state eigenvalue from Eq. \ref{eq:exciton_wannier_equation}.} All projected densities can be defined as
$|\tilde{\psi}_\text{x}(\mathbf{r}_x)|^2=\int \text{d}r_{e1}^2 \text{d}r_{e2}^2 \text{d}r_h^2\left|\psi_T\left(\mathbf{r}_{e1}, \mathbf{r}_{e2}, \mathbf{r}_{h}\right)\right|^2 \delta\left(\mathbf{f}_x\left(\mathbf{r}_{e1}, \mathbf{r}_{e2}, \mathbf{r}_{h}\right) - \mathbf{r}_x\right),$
where $\mathbf{f}_x\left(\mathbf{r}_{e1}, \mathbf{r}_{e2}, \mathbf{r}_{h}\right)$ calculates a transformed coordinate $\mathbf{r}_x$ from the old coordinates, the coordinate transformations are $\mathbf{f}_\text{e}=\mathbf{r}_{e1}$ (\textit{Electrons}), $\mathbf{f}_\text{h}=\mathbf{r}_{h}$ (\textit{Hole}), $\mathbf{f}_\text{COM}=\left(m_e\mathbf{r}_{e1} + m_e\mathbf{r}_{e2} + m_h\mathbf{r}_{h}\right)/M$ (\textit{COM}), $\mathbf{f}_{r_{ee}}=\mathbf{r}_{e2}-\mathbf{r}_{e1}$ (\textit{REL e-e}), and $\mathbf{f}_{r_{eh}}=\mathbf{r}_{h}-\mathbf{r}_{e1}$ (\textit{REL e-h}). In other words, $|\tilde{\psi}_\text{x}(\mathbf{r}_x)|^2$ is the probability distribution of the transformed coordinate $\mathbf{r}_x=\mathbf{f}_x\left(\mathbf{r}_{e1}, \mathbf{r}_{e2}, \mathbf{r}_{h}\right)$.

Additionally, the tables contain the integrand $\chi(\mathbf{r})\coloneq\int \text{d}r_e^2 \psi_T^\ast(\mathbf{r}_e, \mathbf{r}, \mathbf{r})\psi_e(\mathbf{r}_e)$ from the outer integral of the oscillator strength Eq. \ref{eq:trion_oscillator_strength} in the column labeled $r_{eh}=0$.

\begin{table}[tbp]
    \centering
    \includegraphics{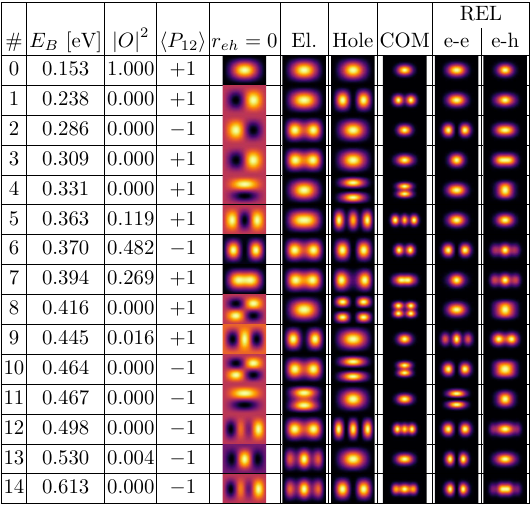}
    \caption{Ground and excited trion states computed for the \npsize{6}{4} nanoplatelet. The columns list the energy $E$ in eV, oscillator strength $|O|^2$ in arbitrary units (not comparable between platelets), the symmetry of the spatial wavefunction under electron exchange $\braket{P_{12}}$ and various densities and wavefunctions: $r_{eh}=0$ is the integrand $\chi(\mathbf{r})$ of the trion oscillator strength. \textit{El.} for the electron $|\tilde{\psi}_e(\mathbf{r}_e)|^2$, \textit{Hole} and \textit{COM} for the hole $|\tilde{\psi}_h(\mathbf{r}_h)|^2$ and center-of-mass $|\tilde{\psi}_\text{COM}(\mathbf{R})|^2$ densities. \textit{REL e-e} for the relative density between the electrons $|\tilde{\psi}_{r_{ee}}(\mathbf{r})|^2$ and \textit{REL e-h} for the relative density between electron and hole $|\tilde{\psi}_{r_{eh}}(\mathbf{r})|^2$.}
    \label{tab:trion_states_6x4}
\end{table}
\begin{table}[tbp]
    \centering
    \includegraphics{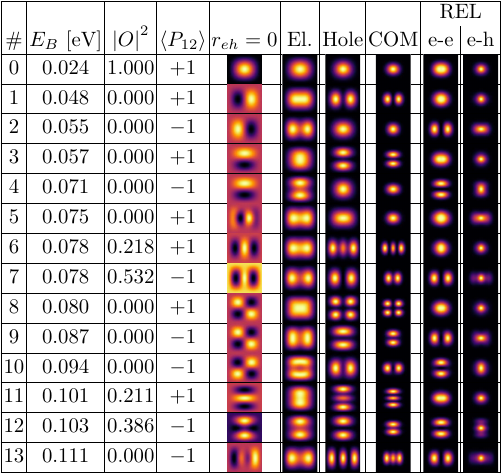}
    \caption{Ground and excited trion states computed for the \npsize{12}{10} nanoplatelet. See Tab. \ref{tab:trion_states_6x4} for explanation of the columns.}
    \label{tab:trion_states_12x10}
\end{table}
\begin{table}[tbp]
    \centering
    \includegraphics[width=\linewidth]{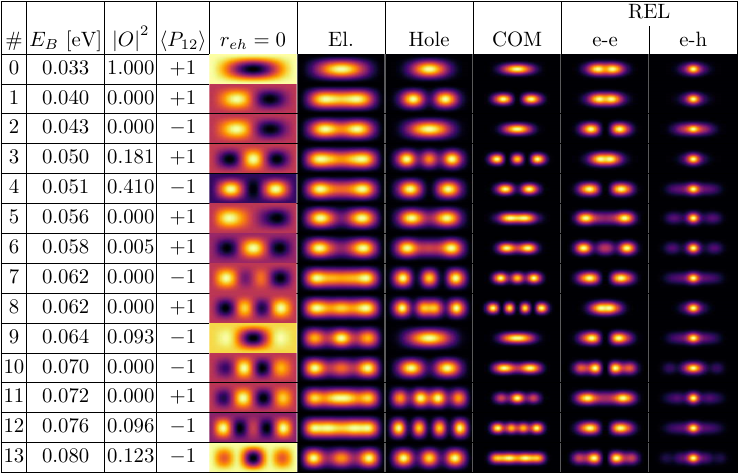}
    \caption{Ground and excited trion states computed for the \npsize{21}{7} nanoplatelet. See Tab. \ref{tab:trion_states_6x4} for explanation of the columns.}
    \label{tab:trion_states_21x7}
\end{table}
\begin{table}[tbp]
    \centering
    \includegraphics{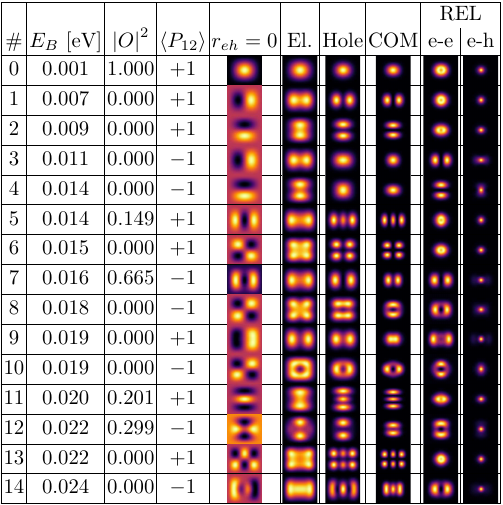}
    \caption{Ground and excited trion states computed for the \npsize{24}{20} nanoplatelet. See Tab. \ref{tab:trion_states_6x4} for explanation of the columns.}
    \label{tab:trion_states_24x20}
\end{table}

\begin{figure}[t]
    \centering
    \includegraphics[width=7.5cm]{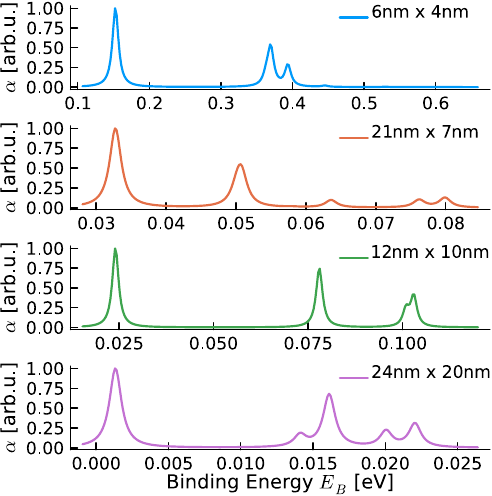}
    \caption{Calculated absorption spectra for the four nanoplatlet sizes. The units are arbitrary and cannot be compared between platelets. The artificial line broadening takes the values $\gamma=5\cdot10^{-3}$ (\npsize{6}{4}), $\gamma=1\cdot10^{-3}$ (\npsize{21}{7}, \npsize{12}{10}) and $\gamma=5\cdot10^{-4}$ (\npsize{24}{20}).}
    \label{fig:trion_absorption_spectra}
\end{figure}

The computed energies and oscillator strengths were also used to calculate artificial absorption spectra using the formula $\alpha(E)=-\text{Im}\left[\sum_n\left|O_n\right|^2/(E_n-E+i\gamma)\right]$ \cite{richter_nanoplatelets_2017}, where $\gamma$ is an artificial line broadening, the results are shown in Fig. \ref{fig:trion_absorption_spectra}.

\subsection{Discussion of the Trion States}
\begin{figure}[h]
    \centering
    \includegraphics[width=6.5cm]{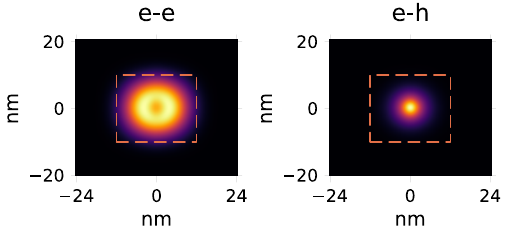}
    \caption{Ground state densities of the relative electron-electron (left) and electron-hole (right) coordinate for the \npsize{24}{20} nanoplatelet. The dashed orange box depicts the size of the nanoplatelet.}
    \label{fig:rel_density_illustration}
\end{figure}

First, we discuss the states of the \npsize{24}{20} nanoplatelet. Here, the low-energy exciton states were well described by the weak-confinement limit. For the trions, however, one can see from the relative coordinate densities that, even in the ground state, the wavefunction almost completely fills the platelet. Note, the plot shows a region twice as large as the platelet (cf. Fig. \ref{fig:rel_density_illustration}). Quantitatively, the average electron-hole distance is $4.7\,\text{nm}$, and the average electron-electron distance is $7\,\text{nm}$. For the exciton, the average electron-hole distance was only $3.6\,\text{nm}$. Additionally, for a \npsize{100}{100} nanoplatelet, the distances in the trion increase further to $10.4\,\text{nm}$ and $6.5\,\text{nm}$, while the distance in the exciton only shows a slight increase to $3.8\,\text{nm}$. It is therefore clear that the trion states of the \npsize{24}{20} platelet are not within the weak-confinement regime.

Another limit would be the strong confinement regime. To test its applicability, we guess the orbitals occurring in the strong confinement ansatz $\psi_T\left(\mathbf{r}_{e1},\mathbf{r}_{e2},\mathbf{r}_h\right)=\left(\phi_{e1}\left(\mathbf{r}_{e1}\right)\phi_{e2}\left(\mathbf{r}_{e2}\right)\pm\phi_{e1}\left(\mathbf{r}_{e2}\right)\phi_{e2}\left(\mathbf{r}_{e1}\right)\right)\phi_{h}\left(\mathbf{r}_{h}\right)/\sqrt{2}$ from the electron and hole densities. With this ansatz, only states in which one electron is in the s-orbital, and the other is in the same orbital as the hole, are bright states.

The bright states of the \npsize{6}{4} nanoplatelet (cf. Tab. \ref{tab:trion_states_6x4}) largely follow the strong confinement expectations: the ground-state densities can be explained by both electrons and the hole being in the s-orbital (s-s-s state); states 6 and 7 can be identified as mostly s-p-p (symmetric and antisymmetric, cf. $\braket{P_{12}}$). The situation for the other bright states is less clear: state 5´s densities indicate an s-s-d state, which should be dark; the hole orbital, which in strong confinement would be a sine with evenly spaced nodes, is modified slightly by the Coulomb interaction, creating a bright state, due to a slight Coulomb-induced asymmetry between negative and positive areas in $r_{eh}=0$. The same is true for state 13, though here the oscillator strength is very small, as for higher-energy states, confinement again starts to dominate over Coulomb. For state 9, the densities clearly indicate a p-p-s state, but this neither matches the integrand of the oscillator strength (a product of p and s has a single node), nor should it be a bright state. The densities and $r_{eh}=0$ of the dark states align well with expectations for strong confinement.  

Similar observations can also be made for the \npsize{12}{10} platelet: the bright states 0, 7, and 12 match the strong confinement ansatz. Whereas bright states 6 and 11 should be dark without the effect of Coulomb interaction, again, they involve a modified d-orbital. 

This phenomenon continues with states 3 (also involving a d-orbital) and 12 of the \npsize{21}{7} platelet. Though for this size, the states 0, 4, 6, and 13 are correctly predicted to be bright by the strong confinement ansatz. For many states, the strong confinement orbitals are hard to guess and do not match the wavefunction overlap $r_{eh}=0$ well. For example, states 5 and 6 have very similar electron and hole densities, but their wave-function overlap, $r_{eh}=0$, is completely different. This indicates that the strong confinement ansatz is no longer suitable to describe the states.

\changes{The Coulomb-brightened states are not only robust against changes of the size and aspect ratio. But, according to tests performed with state 5 of the \npsize{6}{4} nanoplatelet, also persist when the Coulomb potential is modified: by the change to the full Keldysh potential, by varying the dielectric constant of the solvent $\epsilon_{r,out}$ in the tested range from $\epsilon_{r,out}=2$ to $\epsilon_{r,out}=5$, or by the introduction of an exponential screening term $e^{-\rho/\lambda}$ with screening lengths of $\lambda=2\,\text{nm}$ and $\lambda=4\,\text{nm}$.}

\changes{The discrepancies between the strong confinement predictions and the observed states} continue for the \npsize{24}{20} nanoplatelet. Here, only for the ground state, state 3, and state 4, can the observables be satisfactorily explained using the strong confinement ansatz. For many of the other states, the orbitals suggested by the electron and hole densities do not match the forms appearing in $r_{eh}=0$ and thus $\chi(\mathbf{r})$ and the oscillator strength. For example, the densities of state 1 suggest an s-p-p state, which neither matches the wavefunction overlap $r_{eh}=0$ nor the vanishing oscillator strength.

In conclusion, the trions in the two smallest platelets \npsize{6}{4} and \npsize{12}{10} can be described relatively well by the strong confinement ansatz, though some bright states arise due to modifications induced by the Coulomb interaction. It fails, however, for the bigger \npsize{21}{7} and \npsize{24}{20} platelets. Because the average distances between particles are not significantly shorter than any of the platelet sizes studied, we also do not expect the weak-confinement approximation to yield accurate results.

\begin{changesenv}
\subsection{Analysis of required link dimension}
\begin{figure}[t]
    \centering
    \includegraphics[width=\linewidth]{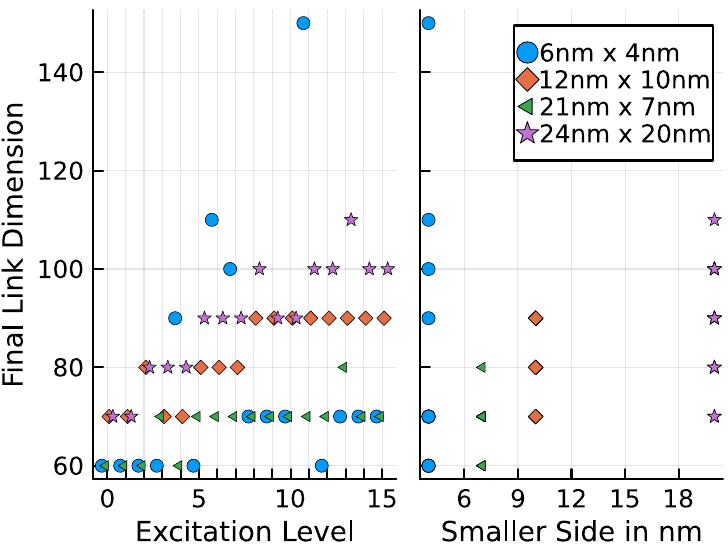}
    \caption{Dependence of the link dimension required for a converged exciton state (resolution $N=11$) on the excitation level and the length of the platelet's smaller dimension. To avoid overlapping points in the excitation level plot the x-axis is slightly shifted for each platelet size.}
    \label{fig:maxdim_vs_exciton_excitation_level_and_smaller_side}
\end{figure}

If the control protocol from Sec. \ref{sec:control_protocol} is used with a constant cutoff of zero and an increasing maximal link dimension, the final link dimension required to reach convergence can be used as a measure of the state's representational complexity in the QTT format. Analysis of these values for the excitons with respect to various parameters shows a slight correlation with the excitation level (Fig. \ref{fig:maxdim_vs_exciton_excitation_level_and_smaller_side}): Especially, for the \npsize{24}{20} and \npsize{12}{10} platelets, the link dimension increases with increasing excitation level. There is also clearly a correlation with the system size, especially with the length of the smaller platelet dimension, where an increase of the link dimension with increasing length is visible in Fig. \ref{fig:maxdim_vs_exciton_excitation_level_and_smaller_side}. We note however, that during the simulation of much
larger platelets, such as the \npsize{100}{100} platelet, we observed a decrease of the required bond dimension.

The trion states require higher link dimensions ranging from 200 to 600 compared to the range of 60 to 150 for the excitons. As for the excitons states, there is a clear increase of the link dimension with increasing smaller platelet dimension.  However, the link dimension seems to stay mostly constant with increasing excitation level: for the \npsize{6}{4} there is a very slight increase, for the \npsize{21}{7} platelet there even is a decrease, and for the \npsize{12}{10} platelet the link dimension for increase and decreases. 

\end{changesenv}

\begin{figure}[t]
    \centering
    \includegraphics[width=\linewidth]{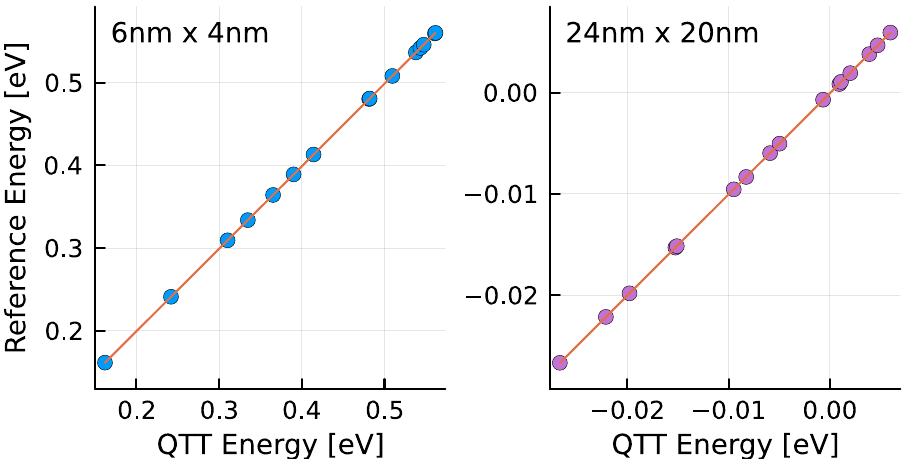}
    \caption{Comparison of energies reported in Ref. \onlinecite{richter_nanoplatelets_2017} and computed using the QTT method at resolution $N=7$ for the \npsize{6}{4} and \npsize{24}{20} nanoplatelets. The orange line indicates a linear fit to the data, $E_{ref}=0.998E_{QTT}-1.07\cdot10^{-4}\,\text{eV}$ for \npsize{6}{4} (left) and $E_{ref}=0.999E_{QTT}-6.06\cdot10^{-5}\,\text{eV}$ for \npsize{24}{20} (right).}
    \label{fig:energy_comparision}
\end{figure}

\begin{figure}[t]
    \centering
    \includegraphics[width=\linewidth]{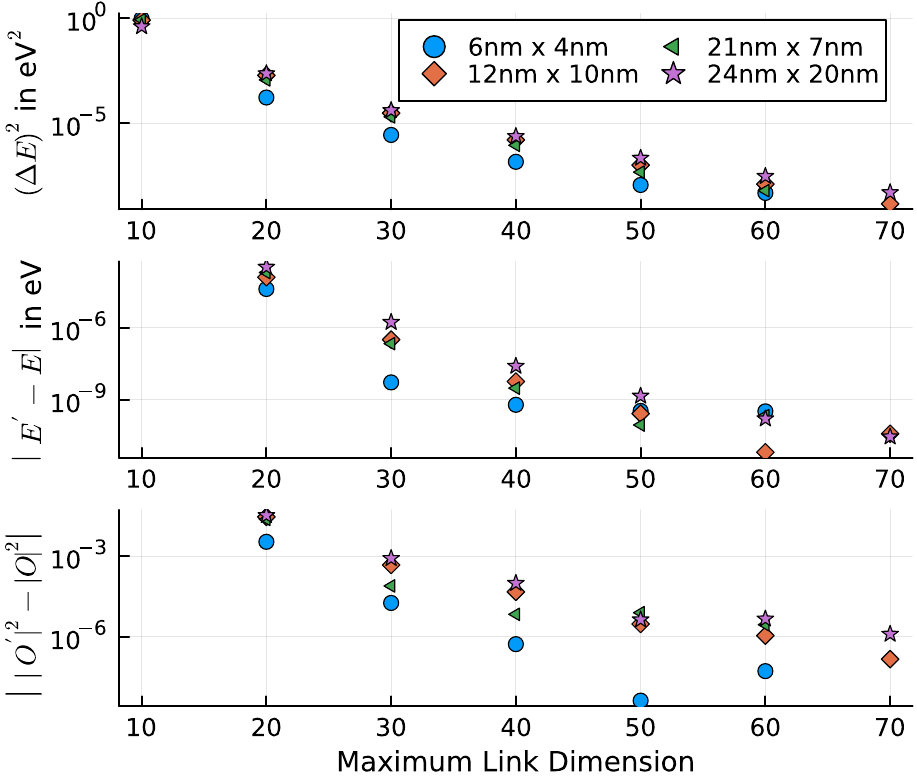}
    \caption{Convergence of energy variance, energy and oscillator strength with respect to the maximum link dimension for the exciton ground states in the different nanoplatelets. For the energy and oscillator strength the difference between the value at the current ($E^\prime$, $\left|O^\prime\right|^2$) and previous ($E$, $\left|O\right|^2$) maximum link dimension is plotted. At resolution $N=11$.}
    \label{fig:convergence_vs_maxdim_exciton_groundstates}
\end{figure}

\section{Conclusion}
In summary, this work demonstrates how tensor networks can be applied to solving multi-particle Schrödinger equations in real space. Specifically, to the calculation of excitons and trions in nanoplatelets without using approximations such as the center-of-mass or relative coordinate factorization, which are only valid in the weak confinement or strong confinement regime, respectively. For both types of quasi-particles, we represented the states in the QTT format and used MPOs that encode binary logical circuits to express the Hamiltonian. The DMRG algorithm was used to obtain ground and excited states with a grid resolution of up to 2048 points per dimension. At this resolution, a calculation using classical direct methods would be infeasible for both excitons and trions. Due to the exponential growth in memory requirements with the number of particles, the trion calculation would be infeasible at any meaningful resolution. Therefore, the calculation of the excitons mainly served to validate the method by comparison with exciton states computed using a direct method.

The calculation of various observables from states expressed in the QTT format was also demonstrated. In some cases, such as the oscillator strength or the electron/hole density, the construction is very simple. Whereas other observables that involve more complicated index transformations, e.g., the center-of-mass density, require more complicated but nonetheless possible tensor network constructions.
We also gained additional insight into the nature of trion states in nanoplatelets, explaining the intricate interplay between strong and weak confinement that differs qualitatively from that of exciton states at typical nanoplatelet sizes. This information was not accessible due to numerical constraints without the tensor network techniques.

\appendix
\section{Comparison with conventional method}
\label{app:conventional_method}
\begin{table}[b]
    \centering
\begin{tabular}{|c|c|c|}
\hline
                & $N=7$ & $N=11$ \\ \hline
\npsize{6}{4}   & $1.095\pm0.29\,\text{meV}$   & $8.237\pm2.152\,\text{meV}$ \\ \hline
\npsize{12}{10} & $0.203\pm0.062\,\text{meV}$  & $1.694\pm0.9\,\text{meV}$   \\ \hline
\npsize{21}{7}  & $0.203\pm0.058\,\text{meV}$  & $1.462\pm0.363\,\text{meV}$ \\ \hline
\npsize{24}{20} & $0.052\pm0.024\,\text{meV}$  & $0.401\pm0.19\,\text{meV}$  \\ \hline
\end{tabular}
    \caption{Mean and standard deviation taken over all matched states of the absolute differences between the energies reported in Ref. 15 and the QTT method for the different platelet sizes at the closest matching resolution $N=7$ and the largest $N=11$.}
    \label{tab:energy_difference_direct}
\end{table}

\changes{Since the reference results \cite{richter_nanoplatelets_2017} were not computed with a power-of-two resolution, they cannot be easily matched for a comparison using the QTT method. Still, even the comparison with mismatched resolutions shows only small errors, the smallest for resolution $N=7$, as listed in Tab. \ref{tab:energy_difference_direct}. The good agreement can also be recognized from Fig. \ref{fig:energy_comparision} where the QTT energies of the \npsize{6}{4} platelet are plotted against the energies of the corresponding states reported in  Ref. \onlinecite{richter_nanoplatelets_2017}. The energies follow a linear relationship very closely, with slope slightly smaller than one, consistent with a slightly smaller resolution for the reference data. (Due to the smaller kinetic energy). The residuals to the proportional fit do not increase with increasing energy, indicating accuracy does not decrease with increasing excitation level using the variance criterion.}

\begin{changesenv}

\section{Convergence analyses}
\begin{figure}[t]
    \centering
    \includegraphics[width=\linewidth]{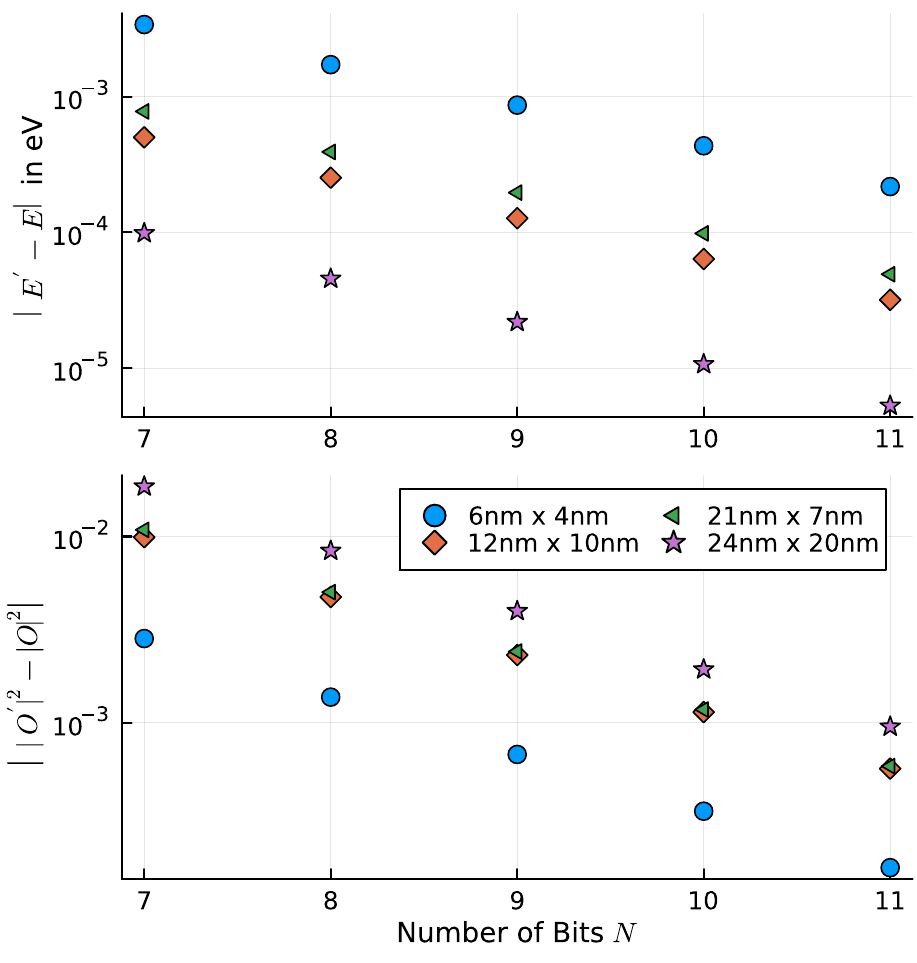}
    \caption{Convergence of energy variance, energy and oscillator strength with respect to the number of bits $N$. For the energy and oscillator strength the difference between the value at the current ($E^\prime$, $\left|O^\prime\right|^2$) and previous ($E$, $\left|O\right|^2$) number of bits is plotted. At variance $(\Delta E)^2\leq 10^{-8}$.}
    \label{fig:convergence_vs_N_exciton_groundstates}
\end{figure}
\label{app:convergence_analyses}
Figure \ref{fig:convergence_vs_maxdim_exciton_groundstates} shows how the energy variance, energy and oscillator strength converge with increasing maximum link dimension based on the example of the exciton ground states. For the energy and oscillator strength the change with regard to the previous lower maximum link dimension is shown, both decrease quickly and are negligibly small ($\leq6.3\cdot10^{-10}\,\text{eV}$ for all exciton states)  earlier than the variance reaches the threshold of $(\Delta E)^2\leq 10^{-8}\,\text{eV}^2$, upon which the computation is stopped. Since the other observables also do not change visibly, we increased the variance threshold slightly to  $(\Delta E)^2\leq 10^{-6}\,\text{eV}^2$ for the trions, which resulted in the energy change at the last truncation to lie below $\leq1.8\cdot10^{-7}\,\text{eV}$ for all trion states. (not shown)

Bigger changes can be seen in Fig. \ref{fig:convergence_vs_N_exciton_groundstates}, which is equivalent to Fig. \ref{fig:convergence_vs_maxdim_exciton_groundstates} except that it shows the convergence with respect to grid resolution not maximum link dimension. Although the changes in oscillator strength and energy are bigger, at $N=11$ both energy and oscillator strength have changes below $10^{-3}$, the last digit reported in the Tables \ref{tab:exciton_state_6x4}-\ref{tab:exciton_state_24x20}. It should also be noted, that the oscillator strength converges more slowly than the energy. For the excited states, which are not shown in the figure, the changes at the last resolution $N=11$ also all lie below $1\,\text{meV}$ and $10^{-3}$ of the ground state's oscillator strength. (Except for in total three states lying at the end of the spectrum, where no equivalent states were calculated during the $N=10$ calculation.)
\end{changesenv}

\begin{changesenv}
\section{Excitons and trions with full Keldysh potential}
\label{app:full_keldysh_excitons}
\begin{figure}[t]
    \centering
    \includegraphics[width=7.5cm]{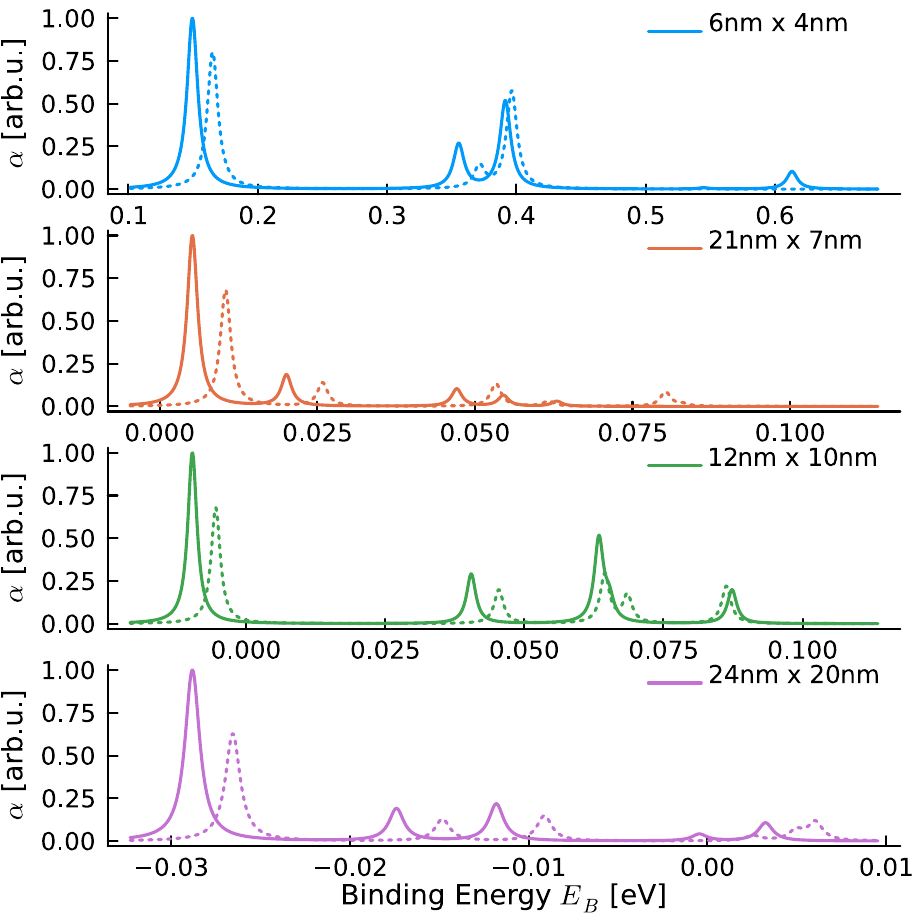}
    \caption{Calculated exciton absorption spectra for the four nanoplatelet sizes. For the solid line the full Keldysh potential was used, for the dashed line the approximate Keldysh potential. The artificial line broadening takes the values $\gamma=5\cdot10^{-3}$ (\npsize{6}{4}), $\gamma=1\cdot10^{-3}$ (\npsize{21}{7}, \npsize{12}{10}) and $\gamma=5\cdot10^{-4}$ (\npsize{24}{20}).}
    \label{fig:full_keldysh_absorption_spectra}
\end{figure}
To test the quality of the approximation of the Keldysh potential the exciton states were also computed using the full Keldysh potential. They are listed in Tables \ref{tab:exciton_full_keldysh_state_6x4} - \ref{tab:exciton_full_keldysh_state_24x20}. 
As can be seen from Fig. \ref{fig:full_keldysh_absorption_spectra}, the resulting absorption spectra are similar, though there seems to be an energy shift and the oscillator strengths are modified.

\begin{figure}[t]
    \centering
    \includegraphics[width=\linewidth]{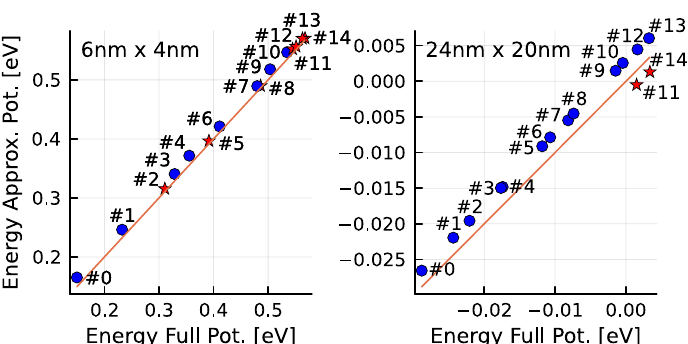}
    \caption{Exciton energies $E_{Approx}$ obtained using the approximate potential over the energies $E_{Keldysh}$ obtained with the full Keldysh potential for similar states with a high overlap. States with a s-like relative density are marked by blue circles, the others by red stars. The orange lines correspond to $E_{Approx}=E_{Keldysh}$. Each state is annotated with the number it takes in the state table Tab. \ref{tab:exciton_full_keldysh_state_6x4} (\npsize{6}{4}) or Tab. \ref{tab:exciton_full_keldysh_state_24x20} (\npsize{24}{20}).}
    \label{fig:energy_comparison_full_keldysh}
\end{figure}

\begin{table}[t]
    \centering
    \includegraphics{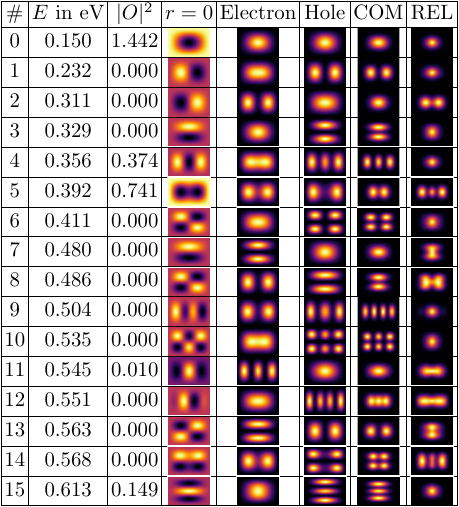}
    \caption{Ground and excited exciton states computed for the \npsize{6}{4} nanoplatelet using the full Keldysh potential. See Tab. \ref{tab:exciton_state_6x4} for explanation of the columns.}
    \label{tab:exciton_full_keldysh_state_6x4}
\end{table}
\begin{table}[t]
    \centering
    \includegraphics{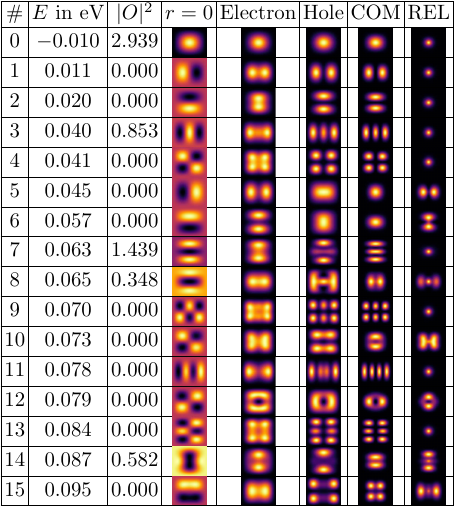}
    \caption{Ground and excited exciton states computed for the \npsize{12}{10} nanoplatelet using the full Keldysh potential. See Tab. \ref{tab:exciton_state_6x4} for explanation of the columns.}
    \label{tab:exciton_full_keldysh_state_12x10}
\end{table}
\begin{table}[t]
    \centering
    \includegraphics[width=\linewidth]{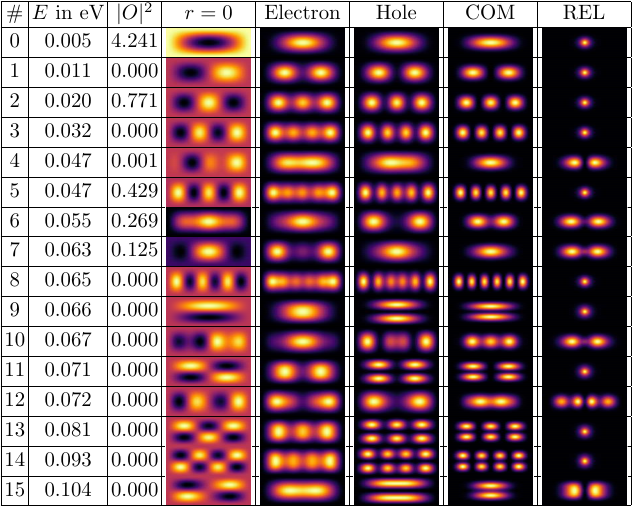}
    \caption{Ground and excited exciton states computed for the \npsize{21}{7} nanoplatelet using the full Keldysh potential. See Tab. \ref{tab:exciton_state_6x4} for explanation of the columns.}
    \label{tab:exciton_full_keldysh_state_21x7}
\end{table}
\begin{table}[t]
    \centering
    \includegraphics{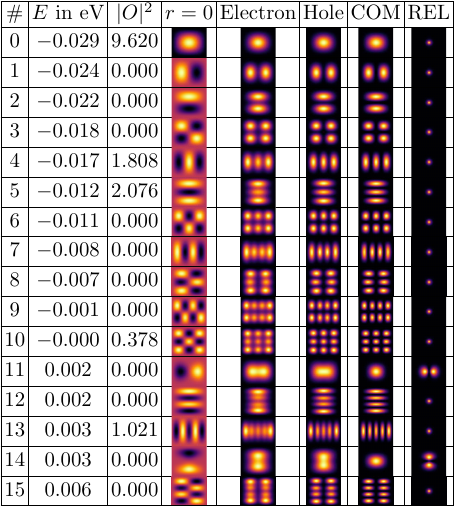}
    \caption{Ground and excited exciton states computed for the \npsize{24}{20} nanoplatelet using the full Keldysh potential. See Tab. \ref{tab:exciton_state_6x4} for explanation of the columns.}
    \label{tab:exciton_full_keldysh_state_24x20}
\end{table}

Figure \ref{fig:energy_comparison_full_keldysh} can be used to compare the energies in more detail: the approximate potential overestimates the energy for most states by an amount that is approximately constant, resulting in a linear relationship between the energies. However, there are a few states where the energy difference is smaller, e.g., states 2 and 5 of the \npsize{6}{4} platelet or states 11 and 14 of the \npsize{24}{20} platelet. These are states where the relative density is not s-like and where the electron and hole are farther apart. This means that the details of the potential near $\mathbf{r}_e-\mathbf{r}_h=0$  are less relevant. Here, the differences between the approximate and full Keldysh potential are the largest. The overestimation of the energy for the s-like states is simply explained by the infinitely deep full potential being approximated by a finitely deep potential. Compared to the energy differences between the states, the errors introduced by the approximate potential (between $~0.015\,\text{eV}$ for the \npsize{6}{4} platelet and $~0.003\,\text{eV}$ for the \npsize{24}{20}) are relatively small compared to typical energies of the respective platelet, but more noticeable for the bigger platelets, where the potential energy is more relevant.

Comparing the oscillator strengths of the ground states, one notices that the approximate Keldysh potential underestimates the oscillator strength, e.g. by factors between $0.80$ (\npsize{6}{4}) and $0.63$ (\npsize{24}{20}) compared to the full Keldysh oscillator strength. This is caused by the stronger close-range attraction, which leads to a smaller exciton Bohr radius, which means that the probability of $\mathbf{r}_e=\mathbf{r}_h$ and thus the amplitude $\psi_{E}(\mathbf{r}, \mathbf{r})$, which appears as the integrand in the exciton oscillator strength, increases. As for the energy, this is only the case for states with s-like relative densities, e.g. state 4 of the \npsize{6}{4}, but not for states with more complicated relative densities, such as state 5 of the same platelet.

There are also some discrepancies between the states computed using the approximate and full Keldysh potential that are simply due to states being missed during the excited states iteration, as explained in Sec. \ref{sec:dmrg_excited_states}, e.g. state 15 from Tab. \ref{tab:exciton_full_keldysh_state_6x4} clearly has no equivalent in \ref{tab:exciton_state_6x4}. To match the states, the overlaps were calculated, and pairs were then selected based on which states had the highest overlap. State pairs with overlap below $80\,\%$ were discarded. Using this method for each platelet the following number of states could be matched out of the 16 that were calculated: 15 (\npsize{6}{4}), 14 (\npsize{21}{7}), 15 (\npsize{12}{10}), and 15 (\npsize{24}{20}). Most states have very high overlap $>98\,\%$.

For the trions only selected states were recomputed using the full Keldysh potential. The energies are again mostly overestimated by the approximate potential with the differences for the ground states lying between $0.020\,\text{eV}$ for the \npsize{6}{4} and $0.003\,\text{eV}$ for \npsize{24}{20}. Likewise, the ground state oscillator strengths are underestimated, with ratios ranging from $0.83$ (\npsize{6}{4}) to $0.67$ (\npsize{24}{20}). These errors are similar to those for the excitons, indicating a similar level of accuracy from the approximate potential.
For the first bright triplet state they are also similar with the energy error between $0.024\,\text{eV}$ (\npsize{6}{4}) and $0.0026\,\text{eV}$ (\npsize{24}{20}) and the oscillator strength underestimated at $0.83$ (\npsize{6}{4}) to $0.58$ (\npsize{24}{20}) of their full Keldysh values. As for the excitons the energy errors are again roughly constant across the states of a single nanoplatelet.
\end{changesenv}

\bibliography{references} 

\end{document}